\documentclass[aps,pre,twocolumn,amssymb]{revtex4-2}

\usepackage{CJK}
\usepackage{graphicx}% Include figure files
\usepackage{dcolumn}% Align table columns on decimal point
\usepackage{bm}% bold math
\usepackage{color}

\usepackage{amsmath}
\usepackage{amssymb}

\def\be{\begin{equation}}
\def\en{\end{equation}}
\def\bea{\begin{eqnarray}}
\def\ena{\end{eqnarray}}
\def\bec{\begin{equation}\begin{array}{rcl}}
\def\p{\partial}

\begin{document}
\begin{CJK*}{UTF8}{}
\title{Reference-Measure Freedom of the Effective Hamiltonian: From Stochastic Thermodynamics to the Macroscopic Limit
  }
\author{Ryuichi Okamoto (\CJKfamily{min}岡本隆一)}
%\email[]{onuki@scphys.kyoto-u.ac.jp}
%\homepage[]{Your web page}
%\thanks{}
%\altaffiliation{}
\affiliation{
 Research Institute for Interdisciplinary Science, Okayama University, Okayama 700-8530, Japan 
}

%Collaboration name if desired (requires use of superscriptaddress
%option in \documentclass). \noaffiliation is required (may also be
%used with the \author command).
%\collaboration can be followed by \email, \homepage, \thanks as well.
%\collaboration{}
%\noaffiliation

%\date{\today}

\begin{abstract}
The effective Hamiltonian (EH), also referred to as the potential of mean
force or free-energy landscape, is widely used to describe equilibrium
properties and slow dynamics in chemical and soft-matter systems.
However, the conventional EH has been argued to be ill-defined because
it does not transform as a scalar under coordinate transformations.
We identify an implicit change of reference measure (gauge), rather
than the coordinate transformation itself, as the origin of this
apparent problem.
We formulate the EH as a scalar relative to a chosen gauge and derive the intrinsic, gauge-independent Fokker--Planck equation (FPE) on the manifold of slow variables. 
The conventional EH and the diffusion-dependent EH introduced in a Riemannian formulation, together with their corresponding FPEs, are thus different gauge representations of the same equilibrium measure and intrinsic stochastic dynamics.
We further demonstrate that the gauge choice has an operational
meaning: different restraint protocols select different reference
measures.
In the macroscopic limit, the leading EH becomes gauge independent
within a class of subleading gauges and is determined by the
large-deviation rate function.
In the same limit, the stochastic dynamics reduces to gauge-independent deterministic dynamics.
\end{abstract}

% insert suggested PACS numbers in braces on next line
%\nopacs{}
% insert suggested keywords - APS authors don't need to do this
%\keywords{}

%\maketitle must follow title, authors, abstract, 
%and \keywords
\maketitle
\end{CJK*}
% body of paper here - Use proper section commands
% References should be done using the \cite, \ref, and \label commands

\section{Introduction} 
The behavior of complex molecular and softmatter systems is often
described in terms of a small number of slow or collective variables.
An effective Hamiltonian (EH), also referred to as the potential of
mean force or free-energy landscape, provides a reduced description of
the equilibrium properties and dynamics of these variables. In principle, it can be obtained computationally from atomistic molecular
simulations \cite{trzesniakComparisonMethodsCompute2007,pietrucciStrategiesExplorationFree2017,segaMolecularUnderstandingFreeEnergy2026}. It can also be reconstructed
from experimental data, particularly in single-molecule measurements
\cite{harrisExperimentalFreeEnergy2007,guptaExperimentalValidationFreeenergylandscape2011,woodsideReconstructingFoldingEnergy2014}.
It has also been constructed from phenomenological consideration for complex systems that are difficult to treat microscopically, yielding great successes in understanding non-trivial behavior in statics and dynamics \cite{Onukibook,DoiBook,Chaikin_Lubensky_1995}. 

Let $\Gamma$ and $H_\mathrm{mic}(\Gamma)$ denote the full-microscopic degrees of freedom and the Hamiltonian of the system, respectively. The EH ${\cal H}({\bm x})$ for a set of chosen variables ${\bm x}$ is usually defined so that $P_\mathrm{eq}({\bm x})d{\bm x}\sim e^{-{\cal H}({\bm x})/T}d{\bm x}$ is the probability measure of ${\bm x}$ in equilibrium, where the Boltzmann constant is set equal to unity and $T$ is the temperature. This definition is expressed as
\begin{align}
  e^{-{\cal H}_{\bm x}({\bm x})/T}=\frac{1}{Z}\int d\Gamma \, e^{-H_\mathrm{mic}(\Gamma)/T}\delta({\bm x}-{\bm X}(\Gamma)). \label{def_H}
\end{align}
There has been a debate that $\cal H$ defined in this manner is not a well-defined scalar with respect to coordinate transformation of ${\bm x}$ and therefore does not provide reliable information on ${\bm x}$ \cite{frenkelSimulationsDarkSide2013,uneyamaDissipationLangevinEquation2020,nakamuraDerivationInvariantFreeEnergy2024}. 

When the dynamics of ${\bm x}$ is substantially slower than the other degrees of freedom, the dynamics of ${\bm x}$ can be described by a white-noise Langevin equation associated with the EH $\cal H$ and the corresponding Fokker-Planck equation (FPE). 
When the diffusion tensor is positive definite, its inverse can naturally be regarded as a Riemannian metric on the manifold ${\cal M}$ of the slow variables ${\bm x}$ \cite{grahamCovariantFormulationNonequilibrium1977}. Based on this geometrical framework, Nakamura recently proposed an expression of the EH that depends on the diffusion matrix and is invariant under coordinate transformation of ${\bm x}$ \cite{nakamuraDerivationInvariantFreeEnergy2024}. 

Now we ask three questions: (i) Is the EH $\cal H$ in Eq.(\ref{def_H}) really physically meaningless? (ii) In softmatter and chemical physics, the EH is often modelled as a function of slow variables or a functional of fields (e.g., a density
field), whose dynamics is described by a stochastic (partial)
differential equation, or by a deterministic one in the macroscopic
limit. If ${\cal H}$ is not a scalar under transformations of these
variables, which variables should be used? (iii) What does it mean that Nakamura's EH depends on the diffusion tensor, which is a dynamical quantity, while the thermodynamic free energy (density) must be free from such a dynamical quantity?

In this paper, we answer these questions by identifying the implicit change of the reference measure underlying the apparent non-scalar transformation of the conventional EH. In fact, the EH is not unique but has a freedom corresponding to the choice of the reference measure; the EH in Eq.(\ref{def_H}) corresponds to the choice of the reference measure as the uniform (Lebesgue) measure $d{\bm x}$, while the Nakamura's EH corresponds to the choice of the Riemannian volume element associated with the diffusion tensor. We show that the FPE satisfying detailed balance with respect to its stationary measure can be formulated in a manner invariant under the choice of reference measure. We further show that a certain class of reference choices has a clear operational meaning and, in this sense, that the EH in Eq. (\ref{def_H}) can indeed be regarded as a physically meaningful scalar once its reference measure is specified.
Finally, we discuss how this reference freedom behaves in the deterministic limit when the equilibrium probability measure has a large deviation form.

\section{Gauge structure of the effective Hamiltonian and stochastic dynamics}
\subsection{Equilibrium measure and gauge freedom}
A probability distribution on the space of slow variables is
intrinsically a measure, rather than a probability density.  A density
is defined only after introducing a reference measure $\nu$, i.e., $\mu = P_\nu \nu$.
This elementary distinction is central to the apparent coordinate
dependence of the EH.  Indeed, if an EH is defined
from a probability density through $P_\nu\propto
e^{-{\cal H}_\nu/T}$, it inevitably requires the choice of reference
measure.  We therefore take the probability measure $\mu$, rather than their corresponding density $P_\nu$ in a particular representation,
as the fundamental, intrinsic object.

Hereafter, we assume that the space of the slow variables is a smooth manifold ${\cal M}$ of dimension $m$. For simplicity, we further assume that ${\cal M}$ is orientable. We thus identify a smooth positive measure with the corresponding positive $m$-form and use the same symbol for both (For non-orientable manifolds, the positive $m$-forms has to be replaced by densities).
Then, for example, we can write $\mu(A)=\int_{X\in A}\mu(X)$ for any measurable set $A\subset {\cal M}$. In local coordinates ${\bm x}=(x^1,\cdots,x^m)$, the $m$-form $\mu$ can be expressed as, for example, $\mu = P_{\bm x}({\bm x}) dx^1 \wedge \cdots \wedge dx^m$. 

Let $X_{\cal M}$ denote the coarse graining map from the microscopic phase space to ${\cal M}$ such that $\Gamma \mapsto X=X_{\cal M}(\Gamma)\in {\cal M}$. 
The equilibrium measure $\mu_\mathrm{eq}$ on ${\cal M}$ is given by the pushforward of the microscopic canonical measure $\mu_\mathrm{mic}=Z^{-1} e^{-H(\Gamma)/T} \nu_\Gamma$ under this map $X_{\cal M}$, where $\nu_\Gamma$ denotes the Liouville measure on the phase space. That is, for any measurable set $A\subset {\cal M}$,
\begin{align}
  \mu_\mathrm{eq}(A)=\mu_\mathrm{mic}(X_{\cal M}^{-1}(A))=\frac{1}{Z}\int_{X_{\cal M}^{-1}(A)} e^{-H(\Gamma)/T} \nu_\Gamma(d\Gamma).
\end{align}
This is a measure-theoretic generalization of Eq.(\ref{def_H}). Choosing a reference measure (positive $m$-form) $\nu$ on ${\cal M}$, we can define the EH ${\cal H}_\nu$ with respect to $\nu$ as
\begin{align}
  e^{-{\cal H}_\nu(X)/T} \nu(X) = \mu_\mathrm{eq} (X), \label{def_H_nu}
\end{align}
where the normalization factor, or equivalently the equilibrium free
energy $F_{\mathrm{eq}}=-T\log Z$, has been absorbed into the definition
of ${\cal H}_\nu$. Equivalently, ${\cal H}_\nu$ is the EH measured relative to the
equilibrium free energy $F_{\mathrm{eq}}=-T\log Z$.
Since the ratio of two $m$-forms $\nu$ and $\mu_\mathrm{eq}$ is a scalar (0-form), the EH ${\cal H}_\nu$ is scalar on ${\cal M}$ once the reference measure $\nu$ is specified.  If we choose the reference measure as the measure $\nu=\nu_{\bm x}\equiv dx^1 \wedge \cdots \wedge dx^m$, which is uniform with respect to the local coordinates ${\bm x}=(x^1,\cdots,x^m)$, we have ${\cal H}_{\nu_{\bm x}} = {\cal H}_{\bm x}$ in Eq.(\ref{def_H}). 

The choice of the reference measure, which is hereafter referred to as the \textit{gauge}, $\nu$ is arbitrary. We can change it to another gauge $\nu'$ as 
\begin{align}
  {\cal H}_{\nu'}(X) = {\cal H}_\nu(X) + T\log \frac{\mathrm{d}\nu'}{\mathrm{d}\nu}(X), \label{gauge_transformation}
\end{align}
where $\mathrm{d}\nu'/\mathrm{d}\nu$ is the Radon-Nikodym derivative of $\nu'$ with respect to $\nu$, which, in this case, expresses just the ``ratio'' of the $m$-forms $\nu'$ and $\nu$ (Eq.~\eqref{def_H_nu} can also be expressed as $e^{-{\cal H}_\nu/T}=\mathrm{d}\mu_\mathrm{eq}/\mathrm{d}\nu$).
Therefore, under the gauge transformation from $\nu_{\bm x}$ to $\nu_{\bm y}=dy^1 \wedge \cdots \wedge dy^m$, which is uniform with respect to another local coordinates ${\bm y}=(y^1,\cdots,y^m)$, the EH transforms as ${\cal H}_{\bm y}({\bm y})={\cal H}_{\bm x}({\bm x}) + T\log \det(\p {\bm y}/\p {\bm x})$, because $\nu_{\bm y}=\det(\p {\bm y}/\p {\bm x}) \nu_{\bm x}$. This relation has been interpreted as the non-scalar transformation of the EH under coordinate transformation \cite{frenkelSimulationsDarkSide2013,uneyamaDissipationLangevinEquation2020,nakamuraDerivationInvariantFreeEnergy2024,yasudaGeometricFormulationStatedependent2026}. In fact, the transformation rule $e^{-{\cal H}_{\bm y}/T} = e^{-{\cal H}_{\bm x}/T}\det(\p {\bm x}/\p {\bm y} ) $ is that of a \textit{density} rather than a scalar. From our viewpoint, however, ${\cal H}_{\bm x}$ and ${\cal H}_{\bm y}$ are both scalar functions corresponding to the two different gauges $\nu_{\bm x}$ and $\nu_{\bm y}$, respectively, and the ``apparent" non-scalar transformation is due to the implicit gauge transformation from $\nu_{\bm x}$ to $\nu_{\bm y}$. 

If the diffusion tensor $D$, which appears in the stochastic dynamics of the slow variables, is a positive-definite contravariant tensor, one may use its inverse to define a Riemannian metric on ${\cal M}$ \cite{grahamCovariantFormulationNonequilibrium1977}. The metric tensor $D^{-1}$ induces its volume $m$-form $\nu_{D}$. In local coordinates ${\bm x}$, it is expressed as $\nu_D=(\det D^{ij})^{-1/2}dx_1 \wedge \cdots \wedge  dx_m=(\det D^{ij})^{-1/2}\nu_{\bm x}$, where $D^{ij}$ is the local coordinate expression of $D$. If we choose $\nu_D$ as the gauge, its associated EH is ${\cal H}_D = {\cal H}_{\bm x}-T\log \sqrt{\det D^{ij}}$, which is indeed the one proposed by Nakamura \cite{nakamuraDerivationInvariantFreeEnergy2024}. Although this choice of Riemannian metric is natural in the mathematics of stochastic thermodynamics, one can also adopt the Riemannian metric of the physical space if the slow variables are the coordinates of the space. The point here is that the equilibrium measure $\mu_\mathrm{eq}$ is the intrinsic physical object, and the pair $(\nu,{\cal H}_\nu)$ provides a representation of it with respect to the chosen reference measure $\nu$. 

\subsection{Operational gauge specification}
The EHs with respect to different gauges are all equivalent representations of $\mu_\mathrm{eq}$. However, some choice of gauges may have a clear operational meaning, which can be used to specify the gauge. Here, we use local coordinates ${\bm x}=(x^1,\cdots,x^m)$, and consider operations of these coordinates by means of external potential forces.

Let us add an restraint potential $kU_\mathrm{ext}({\bm x}(\Gamma),{{\bm \lambda}}(t))$ to the microscopic Hamiltonian $H_\mathrm{mic}(\Gamma)$ in Eq.~\eqref{def_H}, where ${{\bm \lambda}}(t)=(\lambda^1(t),\cdots,\lambda^m(t))$ is a time-dependent external control parameter. We assume that $U_\mathrm{ext}$ has a single minimum at ${\bm x}={{\bm \lambda}}(t)$, and that the minimum value does not depend on ${\bm \lambda}$, i.e.,  $U_\mathrm{ext}({\bm \lambda},{\bm \lambda})=\mathrm{const.}$ The constant $k$ is taken to be sufficiently large to ensure that the slow variables ${\bm x}$ are tightly confined around the minimum of $U_\mathrm{ext}({\bm x},{{\bm \lambda}}(t))$. The work done by the external potential over the time interval $[0,\tau]$ is given by ${\cal W}(\tau)=k\int_0^\tau dt \dot{{\bm \lambda}}(t) \cdot \p_{\bm \lambda} U_\mathrm{ext}$.
The free energy with respect to the total Hamiltonian $H_\mathrm{mic}(\Gamma)+kU_\mathrm{ext}({\bm x},{\bm \lambda})$ is denoted by $F_\mathrm{tot}({\bm \lambda})$. Then, the Jarzynski equality \cite{jarzynskiNonequilibriumEqualityFree1997,jarzynskiEquilibriumFreeenergyDifferences1997} states that 
\begin{align}
  \langle e^{-{\cal W}(\tau)/T} \rangle = e^{- [F_\mathrm{tot}({\bm \lambda}(\tau))-F_\mathrm{tot}({\bm \lambda}(0))]/T}, \label{eq:jarzynski}
\end{align}
where the average $\langle \cdots \rangle$ is taken over the initial microscopic states sampled from the canonical distribution with respect to $H_\mathrm{tot}=H_\mathrm{mic}(\Gamma)+kU_\mathrm{ext}({\bm x}(\Gamma),{\bm \lambda}(0))$. Thus, the Jarzynski equality determines the change in $F_\mathrm{tot}({\bm \lambda})$ along the restraint protocol.
On the other hand, using Eq.~\eqref{def_H} and Laplace's method, $F_\mathrm{tot}({\bm \lambda})$ can be evaluated in the limit
$k\to\infty$ as
\begin{align}
  e^{-F_\mathrm{tot}({\bm \lambda})/T}=&\int d\Gamma\, e^{-H_\mathrm{tot}/T} \nonumber \\
&\hspace{-15mm}=\int d{\bm x} \int d\Gamma\, \delta({\bm x}-{\bm X}(\Gamma)) e^{-(H_\mathrm{mic}(\Gamma)+kU_\mathrm{ext}({\bm x},{\bm \lambda}))/T} \nonumber \\
&\hspace{-15mm}=Z\int d{\bm x}\, e^{-{\cal H}_{\bm x}({\bm x})/T} e^{-kU_\mathrm{ext}({\bm x},{\bm \lambda})/T} \nonumber \\
&\hspace{-15mm}\sim \mathrm{const.}\times\frac{ e^{-{\cal H}_{\bm x}({\bm \lambda})/T} }{\sqrt{\det U_{ij}({\bm \lambda})}} \quad (k\to \infty),
\end{align}
where $U_{ij}({\bm \lambda})=\p_i\p_j U_\mathrm{ext}({\bm x},{\bm \lambda})|_{{\bm x}={\bm \lambda}}$ is the Hessian of $U_\mathrm{ext}$ at the minimum. Therefore, by measuring the work done by the external potential, we can obtain the EH ${\cal H}_{\nu}({\bm x})={\cal H}_{\bm x}({\bm x})+(T/2)\log\det U_{ij}({\bm x})$ with respect to the gauge $\nu=\sqrt{\det U_{ij}({\bm x})} dx^1\wedge \cdots \wedge dx^m$ in local coordinates ${\bm x}$, up to an additive constant (See Eq.~\eqref{gauge_transformation}). This provides a method to operationally specify a gauge.

Therefore, different gauges correspond to different restraint methods represented by the Hessian $U_{ij}({\bm x})$, and are specified by $\nu=\sqrt{\det U_{ij}({\bm x})}\nu_{\bm x}$.  In particular, the standard EH defined in Eq.~\eqref{def_H}, corresponding to the gauge $\nu_{\bm x}$, is a scalar function specified by the operation for $\det U_{ij}=\mathrm{const.}$ This operational construction of the EH has been studied theoretically by Park and Schulten for one-dimensional cases for a large but finite $k$ \cite{park2004}, and also used for calculating the potential of mean force in molecular dynamics simulations \cite{okamoto2024}.
This uniform gauge is also closely related to the free-energy profiles conventionally reconstructed in single-molecule pulling experiments. Hummer and Szabo showed that equilibrium free-energy profiles along a molecular coordinate can be reconstructed from nonequilibrium pulling trajectories using an extension of the Jarzynski equality \cite{hummer2001}, and this approach has subsequently been experimentally validated in single-molecule force-spectroscopy measurements \cite{guptaExperimentalValidationFreeenergylandscape2011}. In the present framework, such a free-energy profile can be interpreted as the EH with respect to the uniform gauge associated with the molecular coordinate used to parameterize the reconstructed profile.

If there is a natural Riemannian metric $g_{ij}$ on the configuration manifold, another natural restraint is to use $g_{ij}$ so that the restraint is compatible with the Riemannian distance. That is, the quadratic-order potential is proportional to the squared Riemannian geodesic distance $d_g({\bm x},{\bm \lambda})^2$, which corresponds to $U_{ij}= \mathrm{const.}\times g_{ij}$ (Conversely, one may use this relation to \textit{define} a metric $g_{ij}$ on ${\cal M}$ from the restraint potential). In this case, the EH with respect to the Riemannian volume measure $\nu_g$ is obtained. As an example, consider the position of a Brownian particle moving on a sphere surface of radius $R$. We choose the Riemannian metric as the standard metric on the sphere, i.e., $g_{\theta\theta}=R^2$, $g_{\phi\phi}=R^2\sin^2\theta$, and $g_{\theta\phi}=g_{\phi\theta}=0$ in the spherical coordinates ${\bm x}=(\theta,\phi)$, which leads to the gauge $\nu_g=R^2\sin\theta\ d\theta\wedge d\phi$. We set $U_\mathrm{ext}\propto d_g(x,\lambda)^2 \approx (\theta -\lambda_\theta)^2g_{\theta\theta}+ (\phi-\lambda_\phi)^2 g_{\phi\phi} = R^2[(\theta -\lambda_\theta)^2+ (\phi-\lambda_\phi)^2\sin^2\lambda_\theta]$. This is a spherically symmetric restraint, and therefore the EH ${\cal H}_{\nu_g}$ with respect to this gauge describes the inhomogeneity of chemical and/or physical properties, if any, encoded in the equilibrium measure $\mu_\mathrm{eq}$, relative to the intrinsic Riemannian volume element.

\subsection{Coordinate- and gauge-invariant Fokker-Planck equation for $\mu_t$}
The EH obtained from simulation, experimental data, or phenomenological consideration, is used to study not only the equilibrium probability density but also the fluctuating or deterministic dynamics of the slow variables. Since the choice of gauge is arbitrary, the physics must not depend on it. 
As in the recent studies \cite{nakamuraDerivationInvariantFreeEnergy2024,yasudaGeometricFormulationStatedependent2026}, let us assume that the dynamics of the slow variables obeys a white-noise Langevin equation. Usually, its corresponding FPE is written in terms of the probability density $P({\bm x},t)$, which already presupposes a reference measure. We instead formulate the evolution directly for the probability
measure $\mu_t$, which makes the coordinate and gauge invariance of the dynamics manifest. A coordinate-free formulation of stochastic processes in terms of differential forms and measures has been developed in Refs.~\cite{PHDThesisBarp,barp2021unifyingcanonicaldescriptionmeasurepreserving}, but this geometric language does not seem to be widely used in stochastic thermodynamics. For reference, Appendix~\ref{app:differential_forms} provides some basic notions and formulas of differential geometry used throughout this paper.

The dynamics on ${\cal M}$ is assumed to obey the stochastic differential equation (SDE) for $X_t\in {\cal M}$, specified in terms of the drift vector field $V$ and the noise vector fields $V_a$. The noise vector fields define the diffusion tensor $D$,
\begin{align}
  D=\frac{1}{2}\sum_a V_a\otimes V_a, \label{eq:D}
\end{align} 
which maps a 1-form $\omega$ to a vector field $D(\omega)=(1/2)\sum_a \omega(V_a)V_a$. 
The SDE is written as
\begin{align}
  {\bf d}X_t = V {\bf d}t + \sum_a V_a \circ {\bf d}W_t^a, \label{SDE}
\end{align}
where $W_t^a$ are independent standard Wiener processes, and $\circ$ denotes the Stratonovich product. 
In the above, we used bold ${\bf d}$ for differentials appearing in SDEs in order to distinguish them from the exterior derivative $d$ on ${\cal M}$. The Radon-Nikodym derivative is denoted by roman $\mathrm{d}$ as in Eq.(\ref{gauge_transformation}). 

The coordinate-free FPE for the measure $\mu_t$ is given by \cite{PHDThesisBarp,barp2021unifyingcanonicaldescriptionmeasurepreserving}
\begin{align}
  \frac{\p \mu_t}{\p t}=-{\cal L}_V \mu_t+\frac{1}{2}\sum_a {\cal L}_{V_a}{\cal L}_{V_a}\mu_t, \label{eq:FPE}
\end{align}
where ${\cal L}_V$ and ${\cal L}_{V_a}$ denote the Lie derivatives along the vector fields $V$ and $V_a$, respectively (See also Appendix~\ref{app:FPE}). Using Cartan's formula ${\cal L}_W=\iota_W d+d\iota_W$ which relates the Lie derivative ${\cal L}_W$ along a vector field $W$, the interior product $\iota_W$, and the exterior derivative $d$, we obtain ${\cal L}_W\mu_t=d\iota_W\mu_t$ as $d\mu_t=0$ \cite{MoritaBook}. Then Eq.~\eqref{eq:FPE} can be rewritten as
\begin{align}
  \frac{\p \mu_t}{\p t}=-d J_\mathrm{irr}(\mu_t), \label{eq:FPE2}
\end{align}
where we have defined a flux $(m-1)$-form $J_\mathrm{irr}(\mu)$ as
\begin{align}
  J_\mathrm{irr}(\mu)=\iota_V\mu-\frac{1}{2}\sum_a \iota_{V_a}{\cal L}_{V_a}\mu \label{eq:flux}
\end{align}
for $m$-form $\mu$. We hereafter assume that the FPE has a stationary (equilibrium) measure $\mu_\mathrm{eq}$ satisfying the detailed-balance condition, 
\begin{align}
  J_\mathrm{irr}(\mu_\mathrm{eq})=\iota _V \mu_\mathrm{eq}-\frac{1}{2}\sum_a \iota_{V_a}{\cal L}_{V_a}\mu_\mathrm{eq}=0. \label{eq:db}
\end{align}
We express $\mu_t$ as $\mu_t=h\mu_\mathrm{eq}$, where $h=\mathrm{d}\mu_t/\mathrm{d}\mu_\mathrm{eq}$ is a scalar.
Substituting $\mu_t=h\mu_\mathrm{eq}$ into Eq.~\eqref{eq:flux} and using Eq.~\eqref{eq:db}, we obtain
\begin{align}
  J_\mathrm{irr}(\mu_t)=-\frac{1}{2}\sum_a[dh(V_a)]\iota_{V_a}\mu_\mathrm{eq},
\end{align}
where we have used ${\cal L}_{V_a}h=dh(V_a)$. We also have $D(dh)=(1/2)\sum_a dh(V_a)V_a$ which yields $\iota_{D(dh)}=(1/2)\sum_a[dh(V_a)]\iota_{V_a}$ (See Eq.~\ref{eq:app_interior_scalar}). We thus have $J_\mathrm{irr}(\mu_t)=-\iota_{D(dh)}\mu_\mathrm{eq}$ to obtain
\begin{align}
  \frac{\p \mu_t}{\p t}=d \left[\iota_{D\left( d\frac{\mathrm{d}\mu_t}{\mathrm{d}\mu_\mathrm{eq}}\right)}\mu_\mathrm{eq}\right]. \label{eq:FPint}
\end{align}
This intrinsic form of the FPE, which is one of the main results of this paper, depends only on the equilibrium measure $\mu_{\rm eq}$ and the diffusion tensor $D$, and is therefore independent of any choice of gauge.

Meanwhile, the map from a vector field to an $(m-1)$-form, $X \mapsto \iota_X \mu_\mathrm{eq}$, is a one-to-one correspondence if we assume $\mu_\mathrm{eq}$ is nonvanishing everywhere on ${\cal M}$. Then, the detailed-balance condition Eq.~\eqref{eq:db} determines the drift vector field $V$ in terms of $\mu_\mathrm{eq}$ and $V_a$. For any vector field $W$ and nonvanishing $m$-form $\mu$, the ratio of the $m$-form ${\cal L}_W\mu$ to $\mu$ defines a scalar function.
The divergence of $W$ with respect to $\mu$, denoted by $\mathrm{div}_\mu W$, is defined as this ratio \cite{LangBook}: 
\begin{align}
  {\cal L}_{W}\mu = (\mathrm{div}_{\mu}W)\mu. \label{eq:div}
\end{align}
Therefore we have ${\cal L}_{V_a}\mu_\mathrm{eq} = (\mathrm{div} _{\mu_\mathrm{eq}}V_a)\mu_\mathrm{eq}$.
Substituting this into Eq.~\eqref{eq:db}, we obtain the  intrinsic expression for the Stratonovich drift vector,
\begin{align}
  V=\frac{1}{2}\sum_a(\mathrm{div}_{\mu_\mathrm{eq}} V_a)V_a. \label{eq:V_intrinsic}
\end{align}
For a specified gauge $\nu$, substituting $\mu_\mathrm{eq}=e^{-{\cal H}_\nu/T}\nu$ into Eq.~\eqref{eq:db}, we obtain gauge-specified expression for $V$,
\begin{align}
  V=-\frac{1}{T} D(d{\cal H}_\nu) +\frac{1}{2}\sum_a \left( \mathrm{div}_\nu V_a\right) V_a.
\end{align}
Here, the first term is the drift induced by the gradient of EH, while the second term represents the noise-induced drift (spurious drift) of Stratonovich-prescription, generalizing the corresponding term derived by Lau and Lubensky for a uniform gauge $\nu=\nu_{\bm x}$ \cite{lau2007}. For a specified gauge $\nu$, defining the probability density $P_\nu$, which is also scalar once gauge $\nu$ is specified, as $\mu_t=P_\nu \nu$, we can rewrite $d(\mathrm{d}\mu_t/\mathrm{d}\mu_\mathrm{eq})=d(e^{{\cal H}_\nu/T}P_\nu)=e^{\cal H_\nu/T}(T^{-1}P_\nu d{\cal H}_\nu +dP_\nu)$ in Eq.~\eqref{eq:FPint}. Therefore we can use \eqref{eq:app_interior_scalar} to rewrite Eq.~\eqref{eq:FPint} as the gauge-specified expression
\begin{align}
  \frac{\p P_\nu}{\p t} = \mathrm{div}_\nu \left[  D(dP_\nu+T^{-1}P_\nu d{\cal H}_\nu)\right]. \label{eq:FPint2}
\end{align}
Here, with the choice of gauge $\nu$, the right-hand side is expressed in terms of the diffusion contribution $\mathrm{div}_\nu(D(dP_\nu))$ and the drift contribution $\mathrm{div}_\nu(DT^{-1}P_\nu d{\cal H}_\nu)$ associated with the EH. It should be noted that the decomposition of the FPE into these two components is itself gauge dependent (while is invariant under coordinate transformations). Only their combination, which determines the evolution of the probability measure \(\mu_t\), is gauge invariant.

We may introduce a mesoscopic internal energy $U_\nu=\int_{\cal M} {\cal H}_\nu \mu_t$ and a mesoscopic entropy $S_\nu=-\int_{\cal M} (\log P_\nu) \mu_t$, which are both gauge dependent. The dependence of entropy on the reference measure has also been discussed in the context of covariant diffusion processes \cite{Polettini_2013}. Their combination, $F_\mathrm{ex}=U_\nu-TS_\nu=T\int_{\cal M} \log (\mathrm{d}\mu_t/\mathrm{d}\mu_\mathrm{eq})\mu_t$, is the free energy excess relative to $F_\mathrm{eq}=-T\log Z$ and is gauge independent. This is $T$ times the Kullback-Leibler divergence of $\mu_t$ from $\mu_\mathrm{eq}$, which is non-negative and monotonically decreasing in time,
\begin{align}
  \frac{dF_\mathrm{ex}}{dt}=-T\int_{\cal M} h^{-2} D(dh,dh)\mu_t \le 0, \label{eq:free_energy_decrease}
\end{align}
where $h=\mathrm{d}\mu_t/\mathrm{d}\mu_\mathrm{eq}$ and $D(dh,dh)=dh(D(dh))=(1/2)\sum_a(dh(V_a))^2\ge0$. The gauge-dependent decomposition of $F_\mathrm{ex}$ into $U_\nu$ and $S_\nu$ is interpreted from the operational viewpoint as follows. We have seen that we can use the Hessian $U_{ij}$ of the restraint potential to specify the gauge, $\nu=\sqrt{\det U_{ij}({\bm x})}dx^1\wedge \cdots \wedge dx^m$. The factor $(\det U_{ij}({\bm x}))^{-1/2}$ represents the position-dependent volume allowed by the potential for the variables to fluctuate, and hence gives an entropic contribution. Therefore, the gauge choice determines how this local entropic contribution is allocated between the internal energy and entropy, while their combination $F_\mathrm{ex}=U_\nu-TS_\nu$ remains gauge independent. The same applies to the gauge-specified expression of FPE in Eq.~\eqref{eq:FPint2}; the decomposition of the intrinsic FPE into diffusion and EH-driven drift is gauge dependent.
The entropy appearing in this decomposition at the mesoscopic level characterizes the distribution of fluctuations over the slow variables and should not, in general, be identified with the macroscopic thermodynamic entropy. Indeed, the EH is not simply an energetic contribution: it also incorporates the entropy of the microscopic degrees of freedom compatible with a given value of the slow variables. When a macroscopic limit exists, as discussed below, the mesoscopic entropy contribution is expected to become subleading at the macroscopic scale, consistently with the disappearance of the diffusive term in the Fokker--Planck equation under the same limit. The EH then converges to the macroscopic thermodynamic free-energy landscape relative to its equilibrium value. In the spirit of the renormalization group, this may be interpreted as follows: the EH absorbs the contribution of the mesoscopic entropy under successive coarse graining, and eventually approaches the macroscopic free-energy landscape.

\subsection{Expressions in local coordinates}
In local coordinates ${\bm x}=(x^1,\cdots,x^m)$, the coordinate basis is given by $\p_i=\p /\p x^i$. The noise vector is expressed as $V_a=V_a^i \partial_i$, using Einstein's summation convention. Substitution of this into Eq.~\eqref{eq:D} yields
\begin{align}
  D=D^{ij}\p_i \otimes \p_j, \quad D^{ij}=\frac{1}{2}\sum_a V_a^iV_a^j. \label{eq:D_local}
\end{align}
The measures $\mu_\mathrm{eq}$ and $\mu_t$ are expressed in terms of scalar functions ${\cal H}_\nu$ and $P_\nu({\bm x},t)$ as
\begin{align}
  \mu_\mathrm{eq}=e^{-{\cal H}_\nu/T}\nu,\quad \mu_t=P_\nu({\bm x},t) \nu.
\end{align}
We can write an arbitrary gauge as
\begin{align}
  \nu=e^{\phi_{\nu,{\bm x}}({\bm x})}dx^1\wedge \cdots \wedge dx^m=e^{\phi_{\nu,{\bm x}}({\bm x})}\nu_{\bm x}, \label{eq:gauge}
\end{align}
with a gauge function $\phi_{\nu,{\bm x}}$. Because the divergence of a vector field $W=W^i\p_i$ is calculated as $\mathrm{div}_\nu W=e^{-\phi_{\nu,{\bm x}}}\p_i (e^{\phi_{\nu,{\bm x}} }W^i)$, the gauge-specified FPE in Eq.~\eqref{eq:FPint2} is expressed in local coordinates as
\begin{align}
  \frac{\p P_\nu}{\p t}=e^{-\phi_{\nu,{\bm x}}}\p_i\left[ e^{\phi_{\nu,{\bm x}} }D^{ij}\left(\p_j P_\nu+T^{-1}P_\nu\p_j{\cal H}_\nu\right)\right]. \label{eq:FPE_local}
\end{align}

Note that the gauge function $\phi_{\nu,{\bm x}}$ is not a scalar under coordinate transformations when the gauge is fixed. To see this, let us perform a coordinate transformation from ${\bm x}$ to ${\bm y}=(y^1,\cdots,y^m)$ while keeping the gauge $\nu$ in Eq.~\eqref{eq:gauge} fixed. The same gauge $\nu$ is expressed as $\nu=e^{\phi_{\nu,{\bm y}}}\nu_{\bm y}=e^{\phi_{\nu,{\bm y}}} J\nu_{\bm x}$ with $J=\det (\p {\bm y}/\p {\bm x})>0$. Comparing this with Eq.~\eqref{eq:gauge}, we obtain
\begin{align}
  \phi_{\nu,{\bm y}}=\phi_{\nu,{\bm x}}-\log J,\quad e^{\phi_{\nu,{\bm y}}}=e^{\phi_{\nu,{\bm x}}} J^{-1}.
\end{align}
This is the transformation rule for the gauge function under the coordinate transformation ${\bm x}\to {\bm y}$. In contrast, by definition, the EH ${\cal H}_\nu$ is a scalar under coordinate transformation with the fixed gauge $\nu$. Under the coordinate transformation ${\bm x}\to {\bm y}$, the EH transforms as a scalar, ${\cal H}_{\nu,{\bm y}}({\bm y})={\cal H}_{\nu,{\bm x}}({\bm x}({\bm y}))$.

The expression of the drift vector in the Stratonovich SDE in Eq.~\eqref{SDE} reads
\begin{align}
  V^i=-\frac{1}{T}D^{ij}\p_j{\cal H}_\nu+\frac{1}{2}\sum_a V_a^i e^{-\phi_{\nu,{\bm x}}} \p_j(e^{\phi_{\nu,{\bm x}}} V_a^j). \label{eq:V_local}
\end{align}
Note that the right hand side does not depend on the choice of the gauge $\nu$ as the intrinsic expression in Eq.~\eqref{eq:V_intrinsic} indicates. It can directly be verified noting that the change $\phi_{\nu,{\bm x}} \to \phi_{\nu,{\bm x}}+\delta\phi$ is accompanied by ${\cal H}_\nu \to {\cal H}_\nu+T\delta\phi$ so that the corresponding changes in the two terms on the right-hand side cancel each other (Similar arguments apply to the FPE Eq.~\eqref{eq:FPE_local}).

For a uniform gauge, i.e., $\phi_{\nu,{\bm x}}=\mathrm{const.}$, the corresponding EH ${\cal H}_\nu$ coincides with ${\cal H}_{\bm x}$ defined in Eq.\eqref{def_H} (apart from a constant shift), and Eqs.~\eqref{eq:FPE_local} and ~\eqref{eq:V_local} reduce to the previous results \cite{lau2007},
\begin{align}
  \frac{\p P_{\bm x}}{\p t}=\p_i\left[ D^{ij}\left(\p_j P_{\bm x}+T^{-1}P_{\bm x}\p_j{\cal H}_{\bm x}\right)\right], \label{eq:FPE_LL}
\end{align}
and
\begin{align}
  V^i=-\frac{1}{T}D^{ij}\p_j{\cal H}_{\bm x}+\frac{1}{2}\sum_a V_a^i \p_j V_a^j. \label{eq:V_LL}
\end{align}

\subsection{Riemannian metric and volume element}
If ${\cal M}$ has a Riemannian metric $g$, we can choose the Riemannian volume element $\nu_g$ as the gauge. A covariant formulation of stochastic
thermodynamics based on such a Riemannian volume measure has been
developed by Ding and Xing \cite{dingCovariantNonequilibriumThermodynamics2022}.
With this choice, the divergence $\mathrm{div}_{\nu_g}$ defined in Eq.\eqref{eq:div} is the standard divergence $\mathrm{div}_g$ with respect to the Riemannian metric $g$ (See Eq.~\eqref{eq:app_divergence}). Furthermore, $g$ induces a natural one-to-one correspondence $\hat g$ between vector fields and 1-forms, such that $\hat g(W) U=g(W,U)$ for any vector fields $W$ and $U$. We also define $\hat D=D \circ \hat g$, i.e., $\hat D(W)=D(\hat g(W))$ for any vector field $W$. Then, the FPE in Eq.~\eqref{eq:FPint2} can be rewritten as
\begin{align}
  \frac{\p P_{\nu_g}}{\p t} = \mathrm{div}_g \left[  \hat D(\mathrm{grad}_g P_{\nu_g}+T^{-1}P_{\nu_g} \mathrm{grad}_g {\cal H}_{\nu_g})\right]. \label{eq:FP_g}
\end{align}
Here, $\mathrm{grad}_g$ denotes the gradient operator with respect to the Riemannian metric $g$ (See Eq.~\eqref{eq:app_gradient}). 

When $D$ is positive-definite, it can be used to define a Riemannian metric such that $D=\hat g^{-1}$ (or $\hat D=1$) \cite{grahamCovariantFormulationNonequilibrium1977,nakamuraDerivationInvariantFreeEnergy2024,yasudaGeometricFormulationStatedependent2026}. Let $\nu_D$ and $P_D$ denote the gauge and probability density such that $\mu_t=P_D \nu_D$, corresponding to this Riemannian metric. This choice simplifies Eq.~\eqref{eq:FP_g} to
\begin{align}
  \frac{\p P_D}{\p t} = \Delta_D P_D+T^{-1}\mathrm{div}_D (P_D \ \mathrm{grad}_D {\cal H}_D), \label{eq:FP_D}
\end{align}
where $\Delta_D=\mathrm{div}_D \mathrm{grad}_D$ is the Laplace-Beltrami operator, and ${\cal H}_D$ is the EH with respect to the gauge $\nu_D$, which is the one proposed by Nakamura \cite{nakamuraDerivationInvariantFreeEnergy2024} as discussed earlier in this paper.
The local coordinate expression of Eqs.~\eqref{eq:FP_g} and \eqref{eq:FP_D} can be obtained using the standard formulas for the divergence and gradient with respect to a Riemannian metric (See Appendix \ref{app:differential_forms}), or, equivalently, by substituting the gauge function $\phi_{\nu,{\bm x}}=\log \sqrt{\det g_{ij}}$ and $\phi_{\nu,{\bm x}}=-\log \sqrt{\det D^{ij}}$, respectively, into Eq.~\eqref{eq:FPE_local}.

The scalar generalized potential introduced in the covariant
stochastic thermodynamics of Ding and Xing~\cite{dingCovariantNonequilibriumThermodynamics2022}
corresponds, in the present terminology, to the representation ${\cal H}_{\nu_g}$ in Eq.~\eqref{eq:FP_g} of the equilibrium measure $\mu_\mathrm{eq}$ with respect to the Riemannian volume form $\nu_g$. The Graham-Nakamura representation, Eq.~\eqref{eq:FP_D}, is mathematically natural as the state dependent diffusion coefficient is absorbed into the differential operators, and, as a result, the diffusion term is simply expressed in terms of the Laplace-Beltrami operator and the drift term is expressed as a gradient of the EH. However, it should be noted that this representation is not unique. The intrinsic FPE in Eq.~\eqref{eq:FPint} is more general and does not require the assumption of a Riemannian metric. It is applicable to any diffusion tensor $D$, including positive-semidefinite and possibly degenerate ones. In such cases, the Graham-Nakamura representation may not be well-defined, but the intrinsic FPE remains valid.

Furthermore, when the slow variables are the coordinates of the physical space, it is more natural to choose the Riemannian metric as the one induced by the physical space. In the case of a Brownian particle on a surface, which was discussed earlier, unless the surface is homogeneous in terms of physical and chemical properties, the EH ${\cal H}_{\nu_g}$ is not a constant, and $D^{-1}$ is not necessarily proportional to $\hat g$. In general, the inhomogeneous stochastic processes has two origins: one is the statics level inhomogeneity, which is reflected in $\mu_\mathrm{eq}$ and hence in the EH ${\cal H}_{\nu_g}$, and the other is the dynamics level inhomogeneity of the diffusion tensor $D$. With the choice of the Riemannian metric as the one induced by the physical space, these two origins of inhomogeneity are clearly separated in the FPE in Eq.~\eqref{eq:FP_g}, which enables a study of the effect of each origin separately. In contrast, they are mixed in the Graham-Nakamura representation in Eq.~\eqref{eq:FP_D}, where ${\cal H}_D$ absorbs the dynamics level inhomogeneity of $D$. Thus, although ${\cal H}_D$ is a valid representation of $\mu_\mathrm{eq}$ with respect to the gauge $\nu_D$, it does not in general represent the static free-energy landscape relative to the physical volume measure.

\subsection{Macroscopic limit and gauge freedom}
We have seen that the EH is a scalar function that has an arbitrariness of gauge transformation, under which, however, the physics is unchanged. The gauge freedom is the freedom of the reference measure, relative to which the equilibrium measure $\mu_\mathrm{eq}$ is expressed. On the other hand, in the macroscopic physics, the thermodynamic free energy and the macroscopic dynamics are deterministic, and contain no explicit probability measure. Then, how does or does not the gauge freedom manifest itself in the macroscopic physics? 

A particularly useful
description of this limit is provided by the large-deviation principle,
which is formulated directly in terms of a family of equilibrium measures
$\mu_{\rm eq}^{(n)}$ \cite{oonoLargeDeviationStatistical1989,touchetteLargeDeviationApproach2009}.  Let $n\to\infty$ characterize the macroscopic limit
and $a_n\to\infty$ denote the corresponding large-deviation speed. For an ordinary thermodynamic limit, $a_n$ may be proportional to
the number of particles, while in a coarse-grained field description its form depends on the scaling between the microscopic and coarse-graining scales.

Suppose that $\mu_{\rm eq}^{(n)}$ satisfies a large-deviation
principle with the rate function $I$.  Formally, for an appropriate set $A\subset{\cal M}$,
this may be written as
\begin{align}
 \mu_{\rm eq}^{(n)}(A)
 \asymp
 \exp\left[
 -a_n\inf_{{\bm x}\in A}I({\bm x})
 \right],
 \label{eq:LDP}
\end{align}
where $\asymp$ denotes equality at the leading exponential order; crudely speaking, it means $a_n^{-1}\log\mu_{\rm eq}^{(n)}(A)\to-\inf_{{\bm x}\in A}I({\bm x})$ as $n\to\infty$ (More precise definition is given in Refs.~\cite{oonoLargeDeviationStatistical1989,touchetteLargeDeviationApproach2009}). The rate function $I$, being defined directly from the family of equilibrium measures, is a scalar function that is intrinsic and independent of any choice of gauge. 

We now seek how this intrinsic rate function is related to the
gauge-dependent EH.  With respect to a gauge $\nu_n$,
we can express the equilibrium measure as
\begin{align}
 \mu_{\rm eq}^{(n)} = e^{-{\cal H}_{\nu_n}^{(n)}/T}\nu_n. \label{eq:mu_eq_nu}
\end{align}
Here, the reference measure $\nu_n$ is subleading relative to the large-deviation speed $a_n$, i.e., we assume that $\nu_n$ is written as $\nu_n=e^{\phi_n}\nu_0$ with a measure $\nu_0$ independent of $n$ and a scalar function $\phi_n$ satisfying $\phi_n=o(a_n)$.
We suppose that the EH ${\cal H}_{\nu_n}^{(n)}$ has the asymptotic form
\begin{align}
 {\cal H}_{\nu_n}^{(n)}({\bm x})=a_n f_\mathrm{ex}({\bm x})+o(a_n).
 \label{eq:macro_H}
\end{align}
We then obtain
\begin{align}
 \int_A \, e^{-{\cal H}_{\nu_n}^{(n)}/T+\phi_n}\nu_0 
  =&\int_A \, e^{-a_n f_\mathrm{ex}/T+o(a_n)}\nu_0 \nonumber \\
  \asymp &\exp\left[-a_n\inf_{{\bm x}\in A}f_\mathrm{ex}({\bm x})/T\right],
\end{align}
where in the last line we have used Laplace's method to evaluate the integral in the limit of $a_n\to\infty$. Comparing this with Eq.~\eqref{eq:LDP}, we obtain
\begin{align}
 I({\bm x})=\frac{f_\mathrm{ex}({\bm x})}{T}.
 \label{eq:rate_EH}
\end{align}
Here, the equilibrium value of $f_{\rm ex}$ is zero, i.e., $\inf_{{\bm x}\in{\cal M}} f_{\rm ex}({\bm x})=0$, consistently with the definition of ${\cal H}_{\nu_n}^{(n)}$, in which the equilibrium free energy $F_{\rm eq}^{(n)}$ is already incorporated as the reference level (see the sentences below Eq.~\eqref{def_H_nu}). Accordingly, $I=f_{\rm ex}/T$ satisfies the conventional condition $\inf I=0$. Thus, the leading macroscopic contribution of the EH is directly related to the large-deviation rate function.

The rate function ``derived'' in this manner is independent of the choice of a subleading gauge, as is expected from the intrinsic nature of the large-deviation principle.  To see this, let us consider another subleading gauge related to $\nu_n$ by
\begin{align}
 \nu_n'=e^{\Delta\phi_n}\nu_n .
 \label{eq:macro_gauge}
\end{align}
Accordingly, the corresponding
EH is
\begin{align}
 {\cal H}_{\nu_n'}^{(n)}
 =
 {\cal H}_{\nu_n}^{(n)}
 +T\Delta\phi_n .
\end{align}
Since we have assumed that both $\nu$ and $\nu'$ are subleading gauges, we have
\begin{align}
 \Delta\phi_n=o(a_n).
 \label{eq:subleading_gauge}
\end{align}
We therefore obtain
\begin{align}
 \lim_{n\to\infty}
 \frac{{\cal H}_{\nu_n'}^{(n)}}{a_n}
 =
 \lim_{n\to\infty}
 \frac{{\cal H}_{\nu_n}^{(n)}}{a_n}
 =
 f_\mathrm{ex} .
 \label{eq:macro_gauge_independence}
\end{align}
Thus, although the EH itself depends on the reference measure at finite
scales, its leading macroscopic contribution is invariant within a class
of subleading gauges relative to the large-deviation speed $a_n$. In particular, the Graham-Nakamura representation of the EH, which is defined with respect to the gauge $\nu_D$ associated with the diffusion tensor $D$, is a subleading gauge if the coordinates ${\bm x}$ represent intensive quantities (such as particle densities) defined in a cell. In such cases, the equilibrium measure may obey a large-deviation principle, $\mu_\mathrm{eq}^N\asymp \exp[-N I({\bm x})]$, where $N$ is the number of microscopic degrees of freedom in the cell, and the large-deviation speed is $a_N=N$. The fluctuations of the intensive quantities typically scale as $N^{-1/2}$, implying $D_N=O(N^{-1})$. Hence the Riemannian volume factor associated with $1/\sqrt{\det (D_N)^{ij}}$ grows only algebraically as $N^{m/2}$, so that its contribution to the EH is $O(\log N)=o(a_N)$, which is subleading. This is an expected result, since the thermodynamic free energy should not depend on the kinetic coefficients.

As an example, let us consider a continuous field such as the local
density in fluctuating hydrodynamics. One may temporarily discretize
such a field into $m$ coarse-graining cells. That is, the coordinates
of ${\cal M}$ are the densities in the cells,
${\bm \rho}=\{\rho^1,\cdots,\rho^m\}$. When the coarse-graining cells
are small in size, the density fluctuations in each cell are large,
and the gauge dependence of the finite-scale EH cannot in general be neglected. To obtain a macroscopic, deterministic hydrodynamic description, we take a limit in which the
size of the coarse-graining cells becomes large compared with the
microscopic scale while remaining infinitesimal compared with the
macroscopic scales. Then, the number $N$ of microscopic degrees of
freedom in each coarse-graining cell grows, and the density fluctuations
in each cell become small. For a fixed number $m$ of coarse-graining cells, the equilibrium measure
is then expected to obey a large-deviation principle of the form
\begin{align}
 \mu_{\rm eq}^{(N,m)}(A)
 \asymp
 \exp\left[
 -N\inf_{{\bm \rho}\in A} I_m({\bm \rho})
 \right],
\end{align}
where the large-deviation speed is $a_N=N$. One may subsequently take
the continuum limit $m\to\infty$, in which the discrete densities
${\bm \rho}$ become a continuous density field $\rho({\bm r})$ and
$I_m$ becomes a rate functional $I[\rho]$. The macroscopic
hydrodynamic description is then determined by this large-deviation
rate functional, which is gauge independent.

Finally, the same large-deviation scaling also explains why the
macroscopic dynamics becomes deterministic and gauge independent.
For simplicity, as in the earlier example, let the large-deviation speed be $N$, so that
\begin{align}
 {\cal H}_{\nu_N}^{(N)}=Nf_\mathrm{ex}+o(N),
 \qquad
 D_N=N^{-1}M+o(N^{-1}).
\end{align}
Substitution into the FPE in Eq.~\eqref{eq:FPint2} shows that the
diffusion contribution vanishes as $O(N^{-1})$, whereas the drift
associated with the EH remains finite,
\begin{align}
 D_N\left(T^{-1}d{\cal H}_{\nu_N}^{(N)}\right)
 \longrightarrow
 T^{-1}M(df_\mathrm{ex})=M(dI),
\end{align}
where Eq.~\eqref{eq:rate_EH} has been used.  The resulting deterministic
dynamics is therefore
\begin{align}
 \dot X=-M(dI).
\end{align}
Since $I$ is intrinsic and gauge independent, the macroscopic
deterministic dynamics is also independent of the choice of a
subleading gauge.  Thus, both the thermodynamic free-energy landscape relative to equilibrium
and the deterministic dynamics emerge from the common leading part of
the gauge-dependent EHs.

\section{Conclusion and remarks}
We have formulated an intrinsic and coordinate-free description of the FPE in Eq.~\eqref{eq:FPint} under the detailed-balance condition on a smooth manifold of the slow variables, using differential forms. This FPE is expressed in terms of the equilibrium measure $\mu_\mathrm{eq}$, which is a top form on the manifold, and the diffusion tensor $D$. Once a reference measure (gauge), which is also a positive top form, is chosen, the equilibrium measure can be expressed in terms of a scalar function, the EH, which is defined with respect to the chosen gauge. The FPE can then be expressed in terms of the EH and the diffusion tensor. In particular, when we choose the inverse diffusion tensor as the Riemannian metric \cite{grahamCovariantFormulationNonequilibrium1977} and its Riemanninan volume form as the gauge, the corresponding EH is the one recently proposed by Nakamura \cite{nakamuraDerivationInvariantFreeEnergy2024}. On the other hand, when we choose the gauge uniform with respect to the local coordinates, the corresponding EH is the standard one defined in Eq.~\eqref{def_H}, and the FPE and drift vector $V$ in Eqs.~\eqref{eq:FPE_LL} and \eqref{eq:V_LL}, respectively, coincide with the Lau-Lubensky form \cite{lau2007}.

The choice of gauge is arbitrary, and different gauges correspond to different representations of the same equilibrium measure and the intrinsic FPE, while the physics does not depend on the choice of gauge. 
This arbitrariness of gauges is not merely a mathematical redundancy: a particular gauge can be given an operational meaning. The gauge can be specified by the operational method used to restrain the slow variables, and different gauges correspond to different restraint methods. Furthermore, for certain problems, there may be a better choice of gauge with which the physical/chemical mechanisms of the system behavior are more clearly represented.

Finally, we have shown that when there is a macroscopic limit in which the equilibrium measure obeys a large-deviation principle, the leading macroscopic contribution of the EH, which is represented by the rate function, is independent of the choice of a subleading gauge. This explains why we do not need to care about the gauge freedom for the macroscopic thermodynamic free-energy landscape relative to equilibrium. We have also shown that the same large-deviation limit of the FPE in Eq.~\eqref{eq:FPint2} leads to a macroscopic, deterministic dynamics, which is also independent of the choice of a subleading gauge. 

Now we are ready to summarize the answers to the three questions (i)--(iii) raised in the introduction. (i) The EH in Eq.~\eqref{def_H} is a scalar function that represents the equilibrium measure with respect to the uniform gauge $dx^1\wedge\cdots\wedge dx^m$. It is specified operationally by the choice of the constant restraint potential, i.e., $\det U_{ij}=\mathrm{const.}$. (ii) Although different gauges give different EHs at finite scales,
their differences are subleading in the macroscopic large-deviation
limit within the class of subleading gauges. Consequently, the leading
macroscopic EH and the resulting deterministic dynamics are independent
of this gauge choice. (iii) At the level of stochastic thermodynamics, the gauge choice is arbitrary, and Nakamura's EH corresponds to the gauge associated with the Riemannian volume form of the inverse diffusion tensor. The physics represented by the intrinsic FPE is independent of the choice of gauge. In the macroscopic limit, the contribution of the diffusion tensor in Nakamura's EH becomes subleading, therefore the thermodynamic free energy is free from the diffusion tensor.

We conclude with a final remark. Throughout this paper, we have restricted ourselves to the irreversible part of the dynamics satisfying the detailed-balance condition, under
which the drift is completely determined by the equilibrium measure and the diffusion tensor. More generally, the dynamics may contain an additional reversible drift
$V_{\rm rev}$ that preserves the equilibrium measure,
\begin{align}
 {\cal L}_{V_{\rm rev}}\mu_{\rm eq}=0,
\end{align}
with the appropriate transformation property under time reversal.
In this case, the drift vector is given by
$V=V_{\rm rev}+V_{\rm irr}$, where $V_{\rm irr}$ is the irreversible
drift determined by the equilibrium measure and the diffusion tensor.
Thus, the $V$ considered in the previous sections corresponds to
$V_{\rm irr}$. For the more general dynamics, the total probability
flux is
\begin{align}
 J(\mu_t)
 =
 J_{\rm irr}(\mu_t)+J_{\rm rev}(\mu_t),
 \quad
 J_{\rm rev}(\mu_t)=\iota_{V_{\rm rev}}\mu_t .
\end{align}
Accordingly, the intrinsic FPE is extended as
\begin{align}
 \frac{\p\mu_t}{\p t}
 =
 -{\cal L}_{V_{\rm rev}}\mu_t
 +d\left[
 \iota_{D\left(d\frac{\mathrm{d}\mu_t}
 {\mathrm{d}\mu_{\rm eq}}\right)}
 \mu_{\rm eq}
 \right].
\end{align}
This more general FPE covers a broad class of stochastic dynamics
satisfying detailed balance with respect to the appropriate
time-reversal transformation.
The additional reversible term is itself intrinsic and introduces no new
dependence on the choice of reference measure. Therefore, the presence of
such a reversible drift does not alter the gauge freedom of the EH or the
gauge independence of the underlying dynamics established in this paper.

%%%%%%%%%%%%%%%%%%%%%%%%%%%%%%%%%%%%%%%%%%%%%
\acknowledgments
The author thanks Kento Yasuda, Shigeyuki Komura, and Masao Doi for informative discussions. This work was supported by Japan Society for the Promotion of Science (JSPS) KAKENHI Grant Number 25K00969.

\appendix

\section{Basic relations and useful formulas in differential geometry}
\label{app:differential_forms}

For completeness, we summarize here some basic formulas for
differential forms used in the main text \cite{MoritaBook,LangBook}. As in the main text, we consider a smooth manifold ${\cal M}$ of dimension $m$. 
\subsection{Interior product}
Let $\alpha_1,\cdots,\alpha_k$ be one-forms and $W_1,\cdots,W_k$ be vector fields.  The evaluation of the $k$-form
$\omega=\alpha_1\wedge\cdots\wedge\alpha_k$ on the vector fields $W_1,\ldots,W_k$ is defined by
\begin{align}
 \omega(W_1,\ldots,W_k)=&\frac{1}{k!}\sum_{\sigma\in S_k}\operatorname{sgn}(\sigma)\prod_{i=1}^k\alpha_i(W_{\sigma(i)}) \nonumber\\
=&\frac{1}{k!}\det\left(\alpha_i(W_j)\right)_{i,j=1,\ldots,k},
 \label{eq:app_wedge}
\end{align}
where $S_k$ is the set of all permutations of $\{1,\ldots,k\}$ and $\operatorname{sgn}(\sigma)$ is the sign of the permutation $\sigma$. Note that there is another common convention that does not include the factor $1/k!$ in Eq.~\eqref{eq:app_wedge}. 

The interior product (or contraction) $\iota_W$ maps a $k$-form to a
$(k-1)$-form.  It is defined by
\begin{align}
 (\iota_W\omega)(W_1,\ldots,W_{k-1})
 =
 k\omega(W,W_1,\ldots,W_{k-1}).
 \label{eq:app_interior}
\end{align}
For a zero-form (scalar function) $f$, it is defined as $\iota_W f=0$. In Eq.~\eqref{eq:app_interior}, the factor $k$ is included to be consistent with the convention in Eq.~\eqref{eq:app_wedge}, and it does not appear in the convention that does not include the factor $1/k!$ in Eq.~\eqref{eq:app_wedge}. Provided that either convention is used consistently, this difference does not affect the equations in the main text or the formulas below.

From the definition, we obviously have the following relations for a scalar function $f$, and vector fields $W_1$ and $W_2$, 
\begin{align}
  \begin{split}
    &\iota_W(f \omega)= f \iota_W \omega=\iota_{f W}\omega\\
    &\iota_{W_1+W_2}\omega=\iota_{W_1}\omega+\iota_{W_2}\omega.
  \end{split}
  \label{eq:app_interior_scalar}
\end{align}
The interior product satisfies
\begin{align}
 \iota_W(\alpha\wedge\beta)
 =
 (\iota_W\alpha)\wedge\beta
 +(-1)^k\alpha\wedge(\iota_W\beta),
 \label{eq:app_interior_product}
\end{align}
where $\alpha$ is a $k$-form. Similar formula for the exterior derivative also holds:
\begin{align}
  d(\alpha\wedge\beta)
  =
  d\alpha\wedge\beta
  +(-1)^k\alpha\wedge d\beta.
  \label{eq:app_exterior_product}
\end{align}
Here, the exterior derivative $d$ of $k$-form is defined in local coordinates by
\begin{equation}
d\left(f\,dx^{i_1}\wedge\cdots\wedge dx^{i_k}\right)
=\sum_j(\partial_j f)\,
dx^j\wedge dx^{i_1}\wedge\cdots\wedge dx^{i_k},
\end{equation}
and extended by linearity.

In the main text, interior products are needed only for one-forms and $m$-forms (top forms). For a one form $\alpha$, Eq.~\eqref{eq:app_interior_product} reduces to
\begin{align}
 \iota_W(\alpha)=\alpha(W).
 \label{eq:app_interior_oneform}
\end{align}
When a top form $\mu$ is expressed in local coordinates $(x^1,\cdots,x^m)$ as $\mu=f\,dx^1\wedge\cdots\wedge dx^m$, and a vector field $W$ is expressed as $W=W^i\partial_i$, we use Eq.~\eqref{eq:app_interior_product} to obtain
\begin{align}
 \iota_W\mu=\sum_{i=1}^m(-1)^{i-1}f W^i dx^1\wedge\cdots\wedge \widehat{dx^i} \wedge\cdots\wedge dx^m,
 \label{eq:app_interior_top}
\end{align}
where $\widehat{dx^i}$ indicates that the factor $dx^i$ is omitted from the wedge product.

\subsection{Lie derivative of differential forms, Cartan's formula, and divergence of vector fields}
The Lie derivative ${\cal L}_W$ of a differential form along $W$ is related to the exterior derivative $d$ and the interior product by Cartan's formula,
\begin{align}
 {\cal L}_W=d\iota_W+\iota_W d.
 \label{eq:app_cartan}
\end{align}
For a scalar function $f$, this gives
\begin{align}
 {\cal L}_W f=\iota_W df=df(W)=W(f).
 \label{eq:app_lie_scalar}
\end{align}
Using Eqs.~\eqref{eq:app_interior_product}, \eqref{eq:app_exterior_product}, and \eqref{eq:app_cartan}, we obtain the Leibniz rule for the Lie derivative of a wedge product of forms,
\begin{align}
  {\cal L}_W(\alpha\wedge\beta)
  =({\cal L}_W\alpha)\wedge\beta
  +\alpha\wedge({\cal L}_W\beta).
  \label{eq:app_lie_product_wedge}
\end{align}
Since an $m$-form $\mu$ on an $m$-dimensional manifold is a top
form, its exterior derivative vanishes identically,
\begin{align}
 d\mu=0.
 \label{eq:app_top_closed}
\end{align}
Cartan's formula therefore reduces to
\begin{align}
 {\cal L}_W\mu
 =
 d(\iota_W\mu).
 \label{eq:app_lie_top}
\end{align}
This identity is used in the main text to express the Fokker--Planck
equation as a continuity equation for the probability measure. If we express the $m$-form $\mu$ in local coordinates as $\mu=f\,dx^1\wedge\cdots\wedge dx^m$, and a vector field $W$ as $W=W^i\partial_i$, we can use Eq.~\eqref{eq:app_interior_top} to obtain
\begin{align}
 {\cal L}_W\mu=\partial_i(f W^i)dx^1\wedge\cdots\wedge dx^m
 \label{eq:app_lie_top_local}
\end{align}
Therefore, if $\mu$ is a positive $m$-form, the divergence of a vector field $W$ with respect to $\mu$, which is defined in Eq.~\eqref{eq:div} in the main text, is expressed in local coordinates as
\begin{align}
 \operatorname{div}_{\mu}W
 =
 \frac{1}{f}\partial_i(f W^i).
 \label{eq:app_div_local}
\end{align}
In particular, if we write $f=e^{\phi}$, we can rewrite Eq.~\eqref{eq:app_div_local} as
\begin{align}
 \operatorname{div}_{\mu}W=e^{-\phi}\partial_i\left(e^\phi W^i\right).
 \label{eq:app_div_coordinate}
\end{align}
No Riemannian metric is required for the definition of divergence in Eq.~\eqref{eq:div}. However, if a Riemannian metric $g$ is given, we can choose the Riemannian volume element $\nu_g$ as the reference measure. Then, the divergence $\mathrm{div}_{\nu_g}$ defined in Eq.~\eqref{eq:div} coincides with the standard divergence $\mathrm{div}_g$ with respect to the Riemannian metric $g$, as we shall see below.

\subsection{Riemannian metric, gradient, divergence, and Laplace-Beltrami operator}
The Riemannian metric $g$ induces a one-to-one linear map $\hat g$ between vector fields and 1-forms, such that $\hat g(W) U=g(W,U)$ for any vector fields $W$ and $U$.
The gradient of a scalar function $f$ is defined as the vector field, 
\begin{align}
  \mathrm{grad}_g f=\hat g^{-1}(df). \label{eq:app_gradient}
\end{align}
With a Riemannian metric $g$ and its associated Riemannian volume element $\nu_g$ for an orientable manifold, Hodge star operator $*$ is defined as a linear map from $k$-forms to $(m-k)$-forms. Let $e_1,\cdots,e_m$ be an orthonormal basis of the tangent space at a point $X\in {\cal M}$, and $\theta^i=\hat g(e_i)$ be their counterparts in the cotangent space identified by the metric $g$. Then, $\theta^1,\cdots,\theta^m$ form an orthonormal basis of the cotangent space. Let $\theta^1,\cdots,\theta^k,\theta^{k+1},\cdots,\theta^m$ be an orthonormal basis of the positive orientation of the cotangent space. The Riemannian volume element $\nu_g$ is defined as 
\begin{align}
  \nu_g=\theta^1\wedge\cdots\wedge\theta^m.
\end{align}
The Hodge star operator for the wedge product of the first $k$ basis one-forms is defined as
\begin{align}
  *(\theta^1\wedge\cdots\wedge\theta^k)=\theta^{k+1}\wedge\cdots\wedge\theta^m.
\end{align}
This is extended to arbitrary $k$-forms by linearity, and is extended to the whole manifold by using the orthonormal basis at each point. 
In particular, for a scalar function $f$, we have $*(f)=f\nu_g$ and $*(f\nu_g)=f$. 

The divergence of a vector field $W$ with respect to the Riemannian metric $g$ is defined as
\begin{align}
  \mathrm{div}_g W = *d*(\hat g(W)). 
\end{align}
We can express $W$ in terms of the orthonormal basis as $W=W^i e_i$, and therefore $\hat g(W)=W_i \theta^i$ with $W^i=W_i$. We then obtain
\begin{align}
  *(\hat g(W))=&W_i *(\theta^i)\nonumber \\
  =&W_i(-1)^{i-1}\theta^1\wedge\cdots\wedge \widehat{\theta^i} \wedge\cdots\wedge \theta^m.
\end{align}
On the other hand, using Eq.~(\ref{eq:app_interior_product}), we obtain
\begin{align}
  \iota_W(\nu_g)=&W^i\iota_{ e_i} (\theta^1\wedge\cdots\wedge \theta^m) \nonumber\\
  =&W_i(-1)^{i-1}\theta^1\wedge\cdots\wedge \widehat{\theta^i} \wedge\cdots\wedge \theta^m. \nonumber\\
  =&*(\hat g(W)).
\end{align}
Therefore, we have
\begin{align}
  \mathrm{div}_g W = *d\iota_W(\nu_g)=* ({\cal L}_W \nu_g)=\mathrm{div}_{\nu_g} W, \label{eq:app_divergence}
\end{align}
where we have used Eq.~\eqref{eq:app_lie_top} in the second equality and Eq.~\eqref{eq:div} in the last equality. This shows that the divergence with respect to the Riemannian metric $g$ coincides with the divergence defined by Eq.~\eqref{eq:div} that assumes no Riemannian structure but only the positive $m$-form $\mu$.

On a Riemannian manifold, the Laplacian or the Laplace-Beltrami operator $\Delta$ for a scalar function $f$ is defined as
\begin{align}
  \Delta f = *d*d f.
\end{align}
Using Eqs.~\eqref{eq:app_gradient} and \eqref{eq:app_divergence}, we can rewrite this as
\begin{align}
  \Delta f = *d*\hat g (\hat g^{-1}(d f))=\mathrm{div}_g \mathrm{grad}_g f.
\end{align}

In the local coordinates ${\bm x}=(x^1,\cdots,x^m)$, the Riemannian metric is expressed as 
\begin{align}
  \hat g=g_{ij}dx^i\otimes dx^j, \label{eq:app_metric_local}
\end{align}
where $g_{ij}$ denotes the inner product of the coordinate basis $\p_i$ and $\p_j$ at each point, i.e., $g_{ij}=g(\p_i,\p_j)$. Then, its inverse is expressed as
\begin{align}
  \hat g^{-1}=g^{ij}\p_i\otimes \p_j,
\end{align}
where $g^{ij}$ is the inverse of the matrix $g_{ij}$, i.e., $g^{ik}g_{kj}=\delta^i_j$.
Then, the gradient of a scalar function $f$ is expressed in local coordinates as
\begin{align}
  \mathrm{grad}_g f =\hat g^{-1}((\p_if)dx^{i})= (g^{ij}\p_j f) \p_i.
\end{align}
Using Eqs.~\eqref{eq:D_local} and \eqref{eq:app_metric_local}, the diffusion tensor $\hat D=D\circ \hat g$ is also expressed in local coordinates as
\begin{align}
  \hat D=D\circ \hat g=D^{il}g_{lj}\p_i\otimes dx^j=D^i_j \p_i\otimes dx^j,
\end{align}
where we have defined $D^i_j=D^{il}g_{lj}$.
Therefore for a vector field $W=W^i\p_i$, we have $\hat D(W)=D^i_j W^j \p_i$.

The local orthonormal basis $\theta^i$ of the cotangent space is expressed as $\theta^i=\theta^i_j dx^j$. Conversely, the coordinate basis $dx^i$ is expressed as $dx^i=(\theta^{-1})^i_j \theta^j$. The inner product of $dx^i$ and $dx^j$ at each point in ${\cal M}$ is given by $(dx^i,dx^j)=g(\hat g^{-1}(dx^i),\hat g^{-1}(dx^j))=g^{ki}g^{lj}g(\p_k,\p_l)=g^{ij}$. Therefore, we have
\begin{align}
  g^{ij}=(dx^i,dx^j)=&(\theta^{-1})^i_k(\theta^{-1})^j_l(\theta^k,\theta^l)\nonumber\\
  =&(\theta^{-1})^i_k(\theta^{-1})^j_l\delta^{kl}\nonumber\\
  =&\sum_k (\theta^{-1})^i_k(\theta^{-1})^j_k,
\end{align}
which yields 
\begin{align}
  \det (g_{ij})=(\det \theta^i_j)^2.
\end{align}
The local coordinate expression of the Riemannian volume element $\nu_g$ is given by
\begin{align}
  \nu_g=&\theta^1\wedge\cdots\wedge\theta^m \nonumber \\
  =&\theta^1_{j_1}\cdots \theta^m_{j_m} dx^{j_1}\wedge\cdots\wedge dx^{j_m}\nonumber \\
  =&(\det \theta^i_j)\ dx^1\wedge\cdots\wedge dx^m \nonumber \\
  =&\sqrt{\det (g_{ij})}\ dx^1\wedge\cdots\wedge dx^m.
\end{align}
Noting Eq.~(\ref{eq:app_divergence}), we can use Eq.~\eqref{eq:app_div_coordinate} to express the divergence of a vector field $W=W^i\p_i$. Substituting $e^\phi=\sqrt{\det (g_{ij})}$ into Eq.~\eqref{eq:app_div_coordinate}, we obtain
\begin{align}
  \mathrm{div}_g W = \frac{1}{\sqrt{\det (g_{kl})}}\p_i\left(\sqrt{\det (g_{kl})} W^i\right). \label{eq:app_div_local_g}
\end{align}
Combining this with Eq.~\eqref{eq:app_gradient}, we can express the Laplace-Beltrami operator $\Delta$ in local coordinates as
\begin{align}
  \Delta f = \frac{1}{\sqrt{\det (g_{kl})}}\p_i\left(\sqrt{\det (g_{kl})} g^{ij}\p_j f\right). \label{eq:app_laplace_beltrami}
\end{align}

In Graham's covariant formulation \cite{grahamCovariantFormulationNonequilibrium1977}, the diffusion tensor $D$ is assumed to be positive-definite, and is used to define a Riemannian metric $g$ such that $\hat g^{-1}=D=D^{ij}\p_i\otimes\p_j$. Then, the local coordinate expressions become
\begin{align}
  &g^{ij}=D^{ij} \\
  &\mathrm{grad}_g f = (D^{ij}\p_j f) \p_i \\
  &\nu_g= \frac{1}{\sqrt{\det (D^{ij})}}\ dx^1\wedge\cdots\wedge dx^m \\
  &\mathrm{div}_g W = \sqrt{\det (D^{kl})}\p_i\left(\frac{W^i}{\sqrt{\det (D^{kl})}}\right) \\
  &\Delta_D f = \sqrt{\det (D^{kl})}\p_i\left(\frac{D^{ij}\p_j f}{\sqrt{\det (D^{kl})}} \right)
\end{align}

\section{Derivation of the coordinate-free Fokker--Planck equation}
\label{app:FPE}
In this Appendix, we give a short derivation of the coordinate-free
Fokker--Planck equation in Eq.~\eqref{eq:FPE} from the Stratonovich stochastic differential equation Eq.~\eqref{SDE}.  In particular, we derive
the infinitesimal generator $L$ directly from the short-time increments of
the Stratonovich stochastic differential equation. 

For a smooth (scalar) function $f$, the Stratonovich differential in Eq.~\eqref{SDE} obeys the ordinary chain rule, therefore we have
\begin{equation}
  {\bf d}f(X_t)=V(f)(X_t)\,{\bf d}t+\sum_a V_a(f)(X_t)\circ{\bf d}W_t^a .
  \label{eq:app_stratonovich_chain}
\end{equation}
We define the infinitesimal generator $L$ as
\begin{align}
  Lf(X)
  &=
  \lim_{\Delta t\to0}
  \frac{
    \langle f(X_{t+\Delta t})-f(X_t)\rangle_{X_t=X}
  }{\Delta t} ,
  \label{eq:app_generator_def}
\end{align}
where $\langle\cdots\rangle_{X_t=X}$ denotes the conditional average at fixed $X_t=X$.  
To derive this operator $L$, we consider a short time interval
$[t,t+\Delta t]$, and denote
\begin{align}
  &\Delta W^a=W_{t+\Delta t}^a-W_t^a \\
  &\Delta f=f(X_{t+\Delta t})-f(X_t)  
\end{align}
Since the infinitesimal generator is a local object, we perform
the following short-time calculation in an arbitrary local
coordinate chart around $X_t$.
Over the short interval, the Stratonovich integral is evaluated at the
midpoint to the required order. 
Since the root-mean-square magnitude of $\Delta W^a$ is of order $\Delta t^{1/2}$, terms quadratic
in $\Delta W^a$ must be retained in the short-time expansion up to order $\Delta t$. We can then integrate the Stratonovich differential in Eq.~\eqref{eq:app_stratonovich_chain} over the interval as
\begin{align}
  \Delta f &=V(f)(X_t)\Delta t+\sum_aV_a(f)\left( X_t+\frac{1}{2}\Delta X \right)\Delta W^a\nonumber\\
  &=V(f)(X_t)\Delta t+\sum_a V_a(f)(X_t)\Delta W^a \nonumber\\
  &\quad +\frac{1}{2}\sum_{a,b}V_b\!\left(V_a(f)\right)(X_t)\Delta W^b\Delta W^a,
  \label{eq:app_short_increment}
\end{align}
where the terms of order $o(\Delta t)$ have been neglected, and in the last term it is sufficient to use
\begin{equation}
  \Delta X=\sum_b V_b(X_t)\Delta W^b+O(\Delta t).
\end{equation}
Using
\begin{align}
  &\langle\Delta W^a\rangle_{X_t=X}=0,\\
  &\langle\Delta W^a\Delta W^b\rangle_{X_t=X}
  =\delta^{ab}\Delta t,
\end{align}
we obtain
\begin{align}
  \langle \Delta f\rangle_{X_t=X}&=\left[V(f)+\frac{1}{2}\sum_a V_a\!\left(V_a(f)\right)
  \right](X)\Delta t +o(\Delta t).
\end{align}
Therefore, from the definition Eq.~\eqref{eq:app_generator_def}, we obtain the infinitesimal generator \cite{PHDThesisBarp}
\begin{equation}
  Lf=V(f)+\frac{1}{2}\sum_a V_a\!\left(V_a(f)\right)={\cal L}_V f+\frac{1}{2}\sum_a {\cal L}_{V_a}{\cal L}_{V_a}f ,
  \label{eq:app_generator}
\end{equation}
where in the second equality we have used Eq.~\eqref{eq:app_lie_scalar}.
This expression is intrinsic and requires neither a metric nor a
volume form.

We next derive the corresponding evolution equation for the probability
measure $\mu_t$.  By the definition of the generator, for any smooth function $f$, we have
\begin{align}
  \frac{d}{dt}\int_{\cal M} f\,\mu_t=\int_{\cal M}(Lf)\,\mu_t .
  \label{eq:app_weak_FPE}
\end{align}
From the Leibniz rule in Eq.~\eqref{eq:app_lie_product_wedge}, for a vector field $W$, a function $h$, and a top form $\mu$, we have ${\cal L}_W (h\mu)=({\cal L}_W h)\mu+h{\cal L}_W \mu$. Furthermore, since $h\mu$ is a top form, we can use Eq.~\eqref{eq:app_lie_top} to obtain
\begin{align}
  \int_{\cal M} d \iota_W (h\mu)=\int_{\cal M}({\cal L}_W h)\mu +\int_{\cal M} h{\cal L}_W \mu \label{eq:app_leibniz_int}
\end{align}
Therefore, if ${\cal M}$ has no boundary, or if the boundary term
vanishes under the imposed boundary conditions, Stokes' theorem implies that the left-hand side of Eq.~\eqref{eq:app_leibniz_int} vanishes. We then have
\begin{equation}
  \int_{\cal M}({\cal L}_W h)\mu = -\int_{\cal M}h\,{\cal L}_W\mu .
  \label{eq:app_Lie_adjoint}
\end{equation}
Applying Eq.~\eqref{eq:app_Lie_adjoint}, we find
\begin{align}
  & \int_{\cal M} ({\cal L}_{V}f)\mu_t =-\int_{\cal M}f\,{\cal L}_{V}\mu_t  \\
  &\int_{\cal M} ({\cal L}_{V_a}{\cal L}_{V_a}f)\mu_t =\int_{\cal M}f\,{\cal L}_{V_a}{\cal L}_{V_a}\mu_t .
  \label{eq:app_Lie_adjoint_result}
\end{align}
Substitution of Eq.~\eqref{eq:app_generator} into
Eq.~\eqref{eq:app_weak_FPE} then gives
\begin{align}
  \frac{d}{dt}\int_{\cal M}f\,\mu_t = \int_{\cal M}f \left[-{\cal L}_V\mu_t+\frac{1}{2}\sum_a {\cal L}_{V_a}{\cal L}_{V_a}\mu_t \right].
\end{align}
Since this holds for arbitrary smooth functions $f$, we obtain Eq.~\eqref{eq:FPE} in the main text \cite{barp2021unifyingcanonicaldescriptionmeasurepreserving}. 

\section{Deterministic limit of FPE}
We derive the deterministic limit without imposing a regularity
condition on the probability density $P_{\nu,N}$. Let
$\mu_{t,N}=P_{\nu,N}\nu$ and
$\mu_{\mathrm{eq},N}=e^{-{\cal H}_{\nu,N}/T}\nu$.
Equation~\eqref{eq:FPint} takes the form
\begin{equation}
  \partial_t\mu_{t,N}=-dJ_N,\quad
  J_N=-\iota_{D_N(dP_{\nu,N}
       +T^{-1}P_{\nu,N}d{\cal H}_{\nu,N})}\nu .
  \label{eq:weak_continuity}
\end{equation}
For a smooth function $f$, $d(fJ_N)=df\wedge J_N+f\,dJ_N$ and Stokes' theorem gives
\begin{equation}
  \frac{d}{dt}\int_{\cal M}f\,\mu_{t,N}
  =\int_{\cal M}df\wedge J_N .
  \label{eq:weak_stokes}
\end{equation}
Here we use Eq.~\eqref{eq:app_interior_product} and symmetric property of $D_N$ to obtain
\begin{align}
  df\wedge \iota_{D_N(dP_{\nu,N})}\nu
  &=df(D_N(dP_{\nu,N}))\nu \nonumber\\
  &= dP_{\nu,N}(D_N(df))\nu \nonumber\\
  &=dP_{\nu,N}\wedge (\iota_{D_N(df)}\nu). \label{eq:app_interior_product_df}
\end{align}
We can also use Eq.~\eqref{eq:app_interior_product} to obtain
\begin{align}
  df\wedge \iota_{D_N(d{\cal H}_{\nu,N})}\nu=df(D_N(d{\cal H}_{\nu,N}))\nu.
\end{align}
Applying Stokes' theorem once more to the integration of \eqref{eq:app_interior_product_df} and using the definition of divergence in Eq.~\eqref{eq:div}, we can rewrite the right-hand side of Eq.~\eqref{eq:weak_stokes} as
\begin{equation}
   \int_{\cal M}\left[
     \operatorname{div}_{\nu}\!\bigl(D_N(df)\bigr)
     -T^{-1}df\bigl(D_N(d{\cal H}_{\nu,N})\bigr)
   \right]\mu_{t,N}.
  \label{eq:weak_generator}
\end{equation}
Here the derivatives act on the test function rather than on
$P_{\nu,N}$.

Suppose that, as $N\to\infty$,
$ND_N\to M$ and $N^{-1}{\cal H}_{\nu,N}\to TI$,
with sufficient regularity for
$\operatorname{div}_{\nu}(D_N(df))\to0$ and
$D_N(d{\cal H}_{\nu,N})\to TM(dI)$. Then we obtain the limiting form of Eq.~\eqref{eq:weak_stokes} as
\begin{align}
  \frac{d}{dt}\int_{\cal M}f\,\mu_t=-\int_{\cal M}df\bigl(M(dI)\bigr)\mu_t=-\int_{\cal M}({\cal L}_{M(dI)}f)\mu_t.
\end{align}
This is indeed the same form as Eq.~\eqref{eq:app_weak_FPE} with $V=-M(dI)$ and $V_a=0$, i.e., the deterministic limit.
This indicates that the deterministic trajectory $X_t$ is governed by
\begin{equation}
  \dot X_t=-M(dI)(X_t).
  \label{eq:macroscopic_flow}
\end{equation}
This argument does not require $P_{\nu,N}$ to vary slowly in
${\cal M}$. In particular, it is compatible with a near-equilibrium
measure of large-deviation form
$\mu_t\approx\mu_{\mathrm{eq},N}\asymp e^{-NI}\nu$, for which the drift
and diffusion contributions to the stationary probability
current can be of the same order.

\bibliography{bibFEL}
\end{document}